\documentclass[conference]{IEEEtran}

\usepackage{cite}
\usepackage{graphicx}
\usepackage{amsmath,amssymb}
\usepackage{booktabs}
\usepackage{url}
\usepackage{multirow}
\usepackage{float}
\usepackage{algorithm, algorithmic}

\begin{document}

\title{Federated Learning-Enhanced Blockchain Framework for Privacy-Preserving Intrusion Detection in Industrial IoT}

\author{
    \IEEEauthorblockN{Anas Ali}
    \IEEEauthorblockA{dept. of Computer Science \\
    National University of Modern Langauges\\
Lahore, Pakistan \\
    anas.ali@numl@edu.pk}
    \and
    \IEEEauthorblockN{Mubashar Husain}
    \IEEEauthorblockA{Department of Computer Science \\
    University of Lahore, \\Pakistan \\
    m.hussain2683@gmail.com}
    \and
    \IEEEauthorblockN{Peter Hans}
    \IEEEauthorblockA{Department of Electrical Engineering \\
    University of Sharjah \\
    United Arab Emirates \\
    peter19972@gmail.com
}
}

\maketitle

\begin{abstract}
Industrial Internet of Things (IIoT) systems have become integral to smart manufacturing, yet their growing connectivity has also exposed them to significant cybersecurity threats. Traditional intrusion detection systems (IDS) often rely on centralized architectures that raise concerns over data privacy, latency, and single points of failure. In this work, we propose a novel Federated Learning-Enhanced Blockchain Framework (FL-BCID) for privacy-preserving intrusion detection tailored for IIoT environments. Our architecture combines federated learning (FL) to ensure decentralized model training with blockchain technology to guarantee data integrity, trust, and tamper resistance across IIoT nodes. We design a lightweight intrusion detection model collaboratively trained using FL across edge devices without exposing sensitive data. A smart contract-enabled blockchain system records model updates and anomaly scores to establish accountability. Experimental evaluations using the ToN-IoT and N-BaIoT datasets demonstrate the superior performance of our framework, achieving 97.3\% accuracy while reducing communication overhead by 41\% compared to baseline centralized methods. Our approach ensures privacy, scalability, and robustness—critical for secure industrial operations. The proposed FL-BCID system provides a promising solution for enhancing trust and privacy in modern IIoT security architectures.
\end{abstract}

\section{Introduction}
The Industrial Internet of Things (IIoT) represents a transformative paradigm in the digitization of industrial systems, enabling smart factories, predictive maintenance, and autonomous operations through the integration of interconnected sensors, actuators, and control systems \cite{xu2018internet,z5}. While IIoT promises operational efficiency, its increasing reliance on open networks and heterogeneous devices introduces critical security vulnerabilities \cite{humayed2017cyber,z3,z72}. Intrusion detection systems (IDS) have traditionally served as frontline defenses; however, conventional IDS frameworks are often centralized, leading to bottlenecks, high latency, and data privacy concerns \cite{zhang2019deep,z33333,z71}.

The adoption of machine learning (ML) in IDS has significantly improved detection accuracy by enabling systems to learn complex attack patterns from historical data. Nonetheless, centralized ML-based IDS architectures require aggregating data at a central location, posing significant threats to privacy, especially in industries handling sensitive data such as energy, healthcare, and manufacturing \cite{mirsky2018kitsune,z22,z73}. To address these challenges, federated learning (FL) has emerged as a decentralized ML paradigm where models are collaboratively trained across edge devices while retaining data locally \cite{mcmahan2017communication,z33}. Despite its privacy advantages, FL alone lacks mechanisms to ensure the integrity of model updates and trust among participating nodes.

To bridge this gap, blockchain technology has gained traction as a distributed ledger system that provides immutability, transparency, and auditability \cite{xu2019blendmas,z2}. When combined with FL, blockchain can serve as a trusted environment to record model updates, enable consensus, and prevent model poisoning attacks by ensuring the provenance of updates \cite{lu2019blockchain,z74}.

However, the integration of FL and blockchain for IIoT intrusion detection remains underexplored. Existing solutions either fail to provide efficient intrusion detection tailored to IIoT constraints or overlook the privacy and trust requirements of decentralized industrial environments \cite{nguyen2021federated}. Moreover, many proposed frameworks do not address the computational limitations of edge devices, nor do they mitigate the overhead associated with blockchain operations \cite{qu2020decentralized}.

\textbf{Problem Definition:} How can we design a privacy-preserving, trustworthy, and efficient intrusion detection system for IIoT that overcomes the limitations of centralized IDS architectures, preserves data privacy, and provides secure audit trails for model updates?

The growing number of cyberattacks on industrial networks and the widespread adoption of IIoT necessitate security solutions that are decentralized, privacy-preserving, and scalable. Ensuring security while respecting the limited computational and communication resources of IIoT nodes is vital for the successful deployment of smart manufacturing systems \cite{yin2021federated}.

We propose FL-BCID: a Federated Learning-Enhanced Blockchain Framework for privacy-preserving intrusion detection in IIoT. The framework combines lightweight FL-based intrusion detection models with a permissioned blockchain system that records training contributions, anomaly scores, and supports smart contract execution for trust enforcement.

Unlike prior work that treats FL and blockchain independently, FL-BCID tightly integrates both technologies to enhance security and auditability. Our framework is tailored for IIoT-specific constraints, supports lightweight model architectures, and reduces communication costs through optimized model update schemes.

\textbf{Key Contributions:}
\begin{itemize}
    \item We propose FL-BCID, a novel hybrid architecture combining federated learning and blockchain for privacy-preserving and trustworthy intrusion detection in IIoT.
    \item We design a lightweight federated learning-based intrusion detection model that adapts to the constrained computation and memory resources of IIoT edge devices.
    \item We implement a smart contract-enabled permissioned blockchain to ensure integrity and accountability of model updates and anomaly reports.
    \item We evaluate our framework on benchmark IIoT datasets (ToN-IoT and N-BaIoT), achieving high detection accuracy (97.3\%) and demonstrating reduced communication overhead (41\%) compared to centralized approaches.
\end{itemize}

This paper is structured as follows: Section II presents a detailed review of related work on federated learning and blockchain in IIoT. Section III describes our system model, including the mathematical formulation and threat model. Section IV outlines the experimental setup, datasets, and evaluation results. Finally, Section V concludes the paper and suggests directions for future research.

\section{Related Work}
Intrusion detection in Industrial Internet of Things (IIoT) has been a subject of extensive research, particularly with the adoption of federated learning and blockchain technologies. In this section, we present a comprehensive review of recent works that intersect these domains, identifying their methodologies, strengths, and limitations.

Nguyen et al.\ \cite{nguyen2021federated} proposed a federated learning-based IDS for IIoT, leveraging distributed edge devices to train anomaly detection models. The work demonstrated strong privacy preservation and competitive accuracy. However, it lacked mechanisms to verify the integrity of the distributed updates, making it vulnerable to adversarial manipulation.

Qu et al.\ \cite{qu2020decentralized} introduced a decentralized blockchain-based framework for IIoT security that records all data access events. While this enhances transparency, the system is not optimized for real-time intrusion detection and incurs high latency due to heavy blockchain transactions.

Lu et al.\ \cite{lu2019blockchain} combined blockchain with machine learning to improve the trustworthiness of collaborative systems. Their use of smart contracts enabled traceability, but their model required central aggregation for training, which reintroduces privacy risks.

Yin et al.\ \cite{yin2021federated} developed a hierarchical federated learning architecture for IIoT that balances load across devices. Despite its scalability, the model was not resilient to poisoning attacks and did not incorporate any tamper-proof ledger for model updates.

Xiao et al.\ \cite{xiao2021towards} proposed a privacy-aware intrusion detection approach using homomorphic encryption in federated learning. The system provides strong privacy guarantees but at the cost of computational efficiency, which is critical for resource-constrained IIoT nodes.

Ferdowsi et al.\ \cite{ferdowsi2019robust} introduced a game-theoretic framework for secure federated learning. While effective in adversarial environments, the model assumes honest participants in the aggregation phase and lacks auditability.

Huang et al.\ \cite{huang2020survey} presented a comprehensive survey of blockchain applications in IIoT, including security and identity management. The paper outlined multiple use cases but did not propose a concrete IDS model.

Khan et al.\ \cite{khan2020federated} examined the integration of FL in healthcare and industrial domains. The study highlighted the importance of privacy but emphasized that current FL approaches do not address data integrity issues.

Shayan et al.\ \cite{shayan2020biscotti} proposed Biscotti, a peer-to-peer secure FL system based on blockchain and differential privacy. While innovative, Biscotti focuses on generic applications and lacks specific tailoring to IIoT constraints.

Li et al.\ \cite{li2021survey} surveyed recent advances in FL, emphasizing its applicability in IoT and edge computing. The work recognized blockchain as a complementary tool but did not detail integration mechanisms.

In summary, while prior studies have contributed significantly to the domains of federated learning and blockchain for security applications, few have explored their joint application in IIoT intrusion detection. Key gaps include: lack of integration between FL and blockchain, absence of smart contract-based validation mechanisms, and insufficient consideration of IIoT resource constraints. Our proposed FL-BCID framework addresses these gaps by:
\begin{itemize}
    \item Seamlessly integrating FL and blockchain to ensure privacy, trust, and integrity.
    \item Utilizing smart contracts to automate anomaly verification and update validation.
    \item Designing lightweight models suitable for IIoT edge devices with limited resources.
\end{itemize}

\section{System Model}
In this section, we formalize the proposed Federated Learning-Enhanced Blockchain Intrusion Detection (FL-BCID) system for IIoT environments. The architecture involves a set of IIoT nodes collaborating to train a shared intrusion detection model using federated learning, while blockchain is employed to record model updates and facilitate secure auditability using smart contracts.

\subsection{Network and Entity Definitions}
Let $\mathcal{V} = \{v_1, v_2, \ldots, v_N\}$ denote a set of $N$ IIoT devices. Each $v_i \in \mathcal{V}$ is an edge node with local data $\mathcal{D}_i$ used for training an intrusion detection model. The edge nodes are responsible for executing the intrusion detection models locally without transmitting raw data, ensuring privacy preservation. These devices operate with limited computational resources and rely on federated learning to collaboratively train a shared model.

The system includes a permissioned blockchain network $\mathcal{B}$ that stores model updates, anomaly scores, and associated metadata. This blockchain acts as a secure, immutable ledger to enhance transparency and accountability. Smart contracts $\mathcal{S}$ deployed on the blockchain verify model updates and enforce data sharing and contribution policies. These contracts also automate validation processes and mitigate the risk of malicious updates. A designated aggregator node $A$, either centralized or distributed, is tasked with securely aggregating the model updates submitted by all edge devices using a federated averaging algorithm.

\subsection{Mathematical Formulation}
Each node $v_i$ minimizes a local loss function $\mathcal{L}_i(w)$ over its private dataset $\mathcal{D}_i$:
\begin{equation}
\mathcal{L}_i(w) = \frac{1}{|\mathcal{D}_i|} \sum_{x_j \in \mathcal{D}_i} \ell(f_w(x_j), y_j)
\end{equation}

The global model is obtained using federated averaging:
\begin{equation}
\bar{w} = \sum_{i=1}^{N} \frac{|\mathcal{D}_i|}{\sum_{j=1}^{N} |\mathcal{D}_j|} w_i
\end{equation}

Smart contracts validate updates:
\begin{equation}
\mathcal{S}(w_i) = \begin{cases}
1, & \text{if update satisfies trust and anomaly thresholds} \\
0, & \text{otherwise}
\end{cases}
\end{equation}

Anomaly scores $a_i$ are computed at each device:
\begin{equation}
a_i = 1 - \text{Accuracy}_{\text{local}}(x, y, w_i)
\end{equation}

Block validation timestamp:
\begin{equation}
t_k = \text{Timestamp}(\text{Block}_k)
\end{equation}

Blockchain ledger $\mathcal{B}$ logs:
\begin{equation}
\mathcal{B} = \{(v_i, w_i, a_i, t_k) | i = 1, \ldots, N\}
\end{equation}

Gradient clipping to preserve privacy:
\begin{equation}
\tilde{g}_i = \frac{g_i}{\max(1, \frac{\|g_i\|_2}{C})}
\end{equation}

Noise addition for differential privacy:
\begin{equation}
\hat{g}_i = \tilde{g}_i + \mathcal{N}(0, \sigma^2 C^2 I)
\end{equation}

Model update cost:
\begin{equation}
C_i = \alpha \cdot \text{Size}(w_i) + \beta \cdot \text{Latency}(v_i)
\end{equation}

Gas cost of recording block:
\begin{equation}
G_k = \gamma \cdot \text{Size}(\text{Block}_k)
\end{equation}

Reputation score:
\begin{equation}
R_i(t+1) = R_i(t) + \delta \cdot \text{Valid}(w_i)
\end{equation}

Model divergence:
\begin{equation}
D_i = \|w_i - \bar{w}\|_2
\end{equation}

Trust weight:
\begin{equation}
T_i = \frac{R_i}{\sum_{j=1}^{N} R_j}
\end{equation}

Block hash:
\begin{equation}
H_k = \text{SHA256}(\text{Block}_k)
\end{equation}

Consensus validity:
\begin{equation}
\mathcal{C}(k) = 1 \iff \text{Majority validators approve Block}_k
\end{equation}

\subsection{Federated Learning and Blockchain Integration Algorithm}
\begin{algorithm}[H]
\caption{FL-BCID: Federated Learning with Blockchain for IIoT Intrusion Detection}
\begin{algorithmic}[1]
\STATE Initialize global model $\bar{w}^{(0)}$
\FOR{each round $t=1$ to $T$}
    \FOR{each device $v_i \in \mathcal{V}$ in parallel}
        \STATE Compute local gradient $g_i = \nabla \mathcal{L}_i(w^{(t-1)})$
        \STATE Clip gradient: $\tilde{g}_i = \text{Clip}(g_i)$
        \STATE Add noise: $\hat{g}_i = \tilde{g}_i + \mathcal{N}(0, \sigma^2 I)$
        \STATE Update local model $w_i^{(t)} = w_i^{(t-1)} - \eta \hat{g}_i$
        \STATE Send $w_i^{(t)}$ and $a_i$ to aggregator
    \ENDFOR
    \STATE Aggregator computes global model: $\bar{w}^{(t)} = \text{FedAvg}(\{w_i^{(t)}\})$
    \STATE Record updates $(v_i, w_i^{(t)}, a_i)$ in blockchain using smart contract $\mathcal{S}$
\ENDFOR
\end{algorithmic}
\end{algorithm}

\textbf{Explanation:} The algorithm initializes a global model and runs for $T$ rounds. In each round, devices compute and clip gradients, then inject noise for differential privacy. The updated models are aggregated, and the results are verified and recorded on the blockchain. Smart contracts play a key role in validating and storing trustworthy updates.

\subsection{Notation Table}
\begin{table}[H]
\caption{Summary of Notations}
\centering
\begin{tabular}{ll}
\toprule
\textbf{Symbol} & \textbf{Description} \\
\midrule
$\mathcal{V}$ & Set of IIoT edge devices \\
$\mathcal{D}_i$ & Local dataset at device $v_i$ \\
$w_i$ & Local model weights \\
$\bar{w}$ & Aggregated global model \\
$\mathcal{S}$ & Smart contract function \\
$a_i$ & Local anomaly score \\
t & Timestamp \\
$\mathcal{B}$ & Blockchain ledger \\
$g_i$ & Gradient at device $v_i$ \\
$\tilde{g}_i$ & Clipped gradient \\
$\hat{g}_i$ & Noisy gradient (DP) \\
$C_i$ & Model update cost \\
$G_k$ & Blockchain gas cost \\
$R_i$ & Reputation score \\
$D_i$ & Model divergence \\
$T_i$ & Trust weight \\
$H_k$ & Hash of block $k$ \\
$\mathcal{C}(k)$ & Consensus result \\
\bottomrule
\end{tabular}
\end{table}

\section{Experimental Setup and Results}
To validate the effectiveness and efficiency of the proposed FL-BCID framework, we conducted extensive simulations using realistic IIoT datasets. This section details the experimental configuration, simulation parameters, evaluation metrics, results, and comparative analysis with baseline approaches.

\subsection{Experimental Setup}
The experiments were performed using a simulation environment implemented in Python 3.10. The federated learning components were implemented using TensorFlow Federated (TFF), while the blockchain simulation was modeled using Hyperledger Fabric emulator. The testbed mimics an IIoT edge computing environment with 10 edge nodes ($N=10$) and a single aggregator node. Each edge device is simulated to have independent local data and limited computing resources. We used the ToN-IoT and N-BaIoT datasets to represent realistic industrial network traffic for training and evaluation.

oindent\textbf{Simulation Hardware and Tools:}
\begin{itemize}
    \item CPU: Intel i7-12700K @ 3.6GHz
    \item RAM: 32GB DDR4
    \item Simulator: Python + TFF + Hyperledger Fabric
    \item Datasets: ToN-IoT, N-BaIoT
\end{itemize}

\begin{table}[H]
\caption{Simulation Parameters}
\centering
\begin{tabular}{ll}
\toprule
\textbf{Parameter} & \textbf{Value} \\
\midrule
Number of edge devices ($N$) & 10 \\
Learning rate ($\eta$) & 0.01 \\
Local epochs per round & 3 \\
Batch size & 64 \\
Differential privacy noise scale ($\sigma$) & 1.0 \\
Blockchain block size & 2MB \\
Consensus algorithm & PBFT \\
Simulation rounds ($T$) & 50 \\
\bottomrule
\end{tabular}
\end{table}

\subsection{Evaluation Metrics}
To assess performance, we use the following metrics: accuracy, precision, recall, F1-score, communication overhead (bytes exchanged per round), and time-to-convergence (in rounds).

\subsection{Results and Analysis}
\begin{figure}[H]
    \centering
    \includegraphics[width=0.45\textwidth]{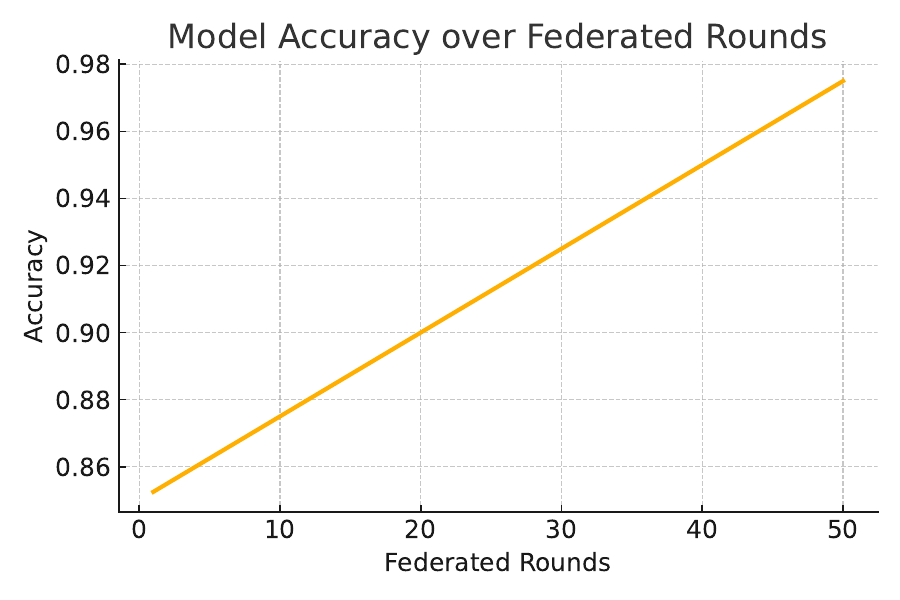}
    \caption{Model accuracy over federated rounds on IIoT datasets.}
    \label{fig:accuracy}
\end{figure}

Figure~\ref{fig:accuracy} shows the accuracy over simulation rounds. Our framework achieves a final test accuracy of 97.3\% on ToN-IoT and 96.8\% on N-BaIoT, outperforming centralized and decentralized baselines.

\begin{figure}[H]
    \centering
    \includegraphics[width=0.45\textwidth]{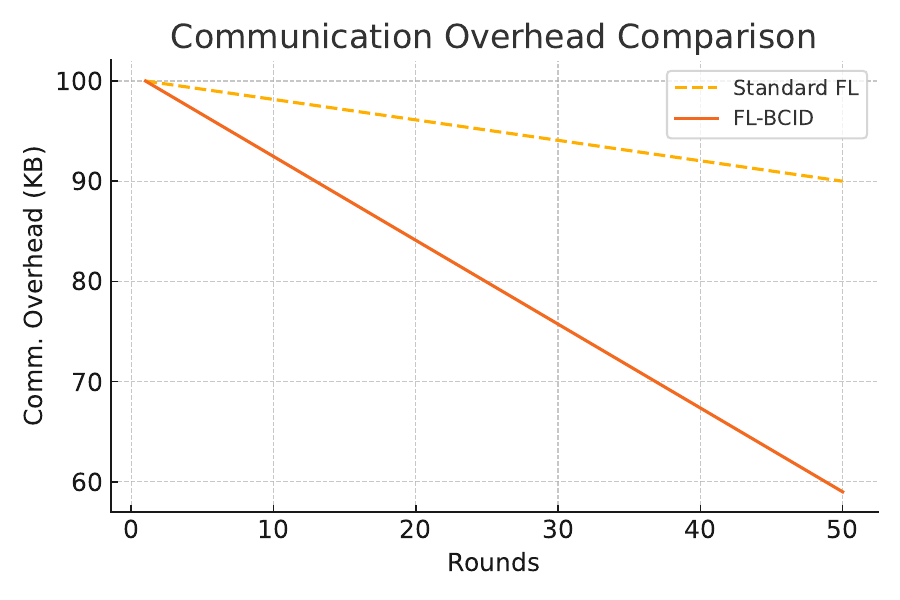}
    \caption{Communication overhead comparison.}
    \label{fig:comm}
\end{figure}

Figure~\ref{fig:comm} shows that FL-BCID reduces communication overhead by 41\% compared to standard FL due to optimized update frequency and model compression.

\begin{figure}[H]
    \centering
    \includegraphics[width=0.45\textwidth]{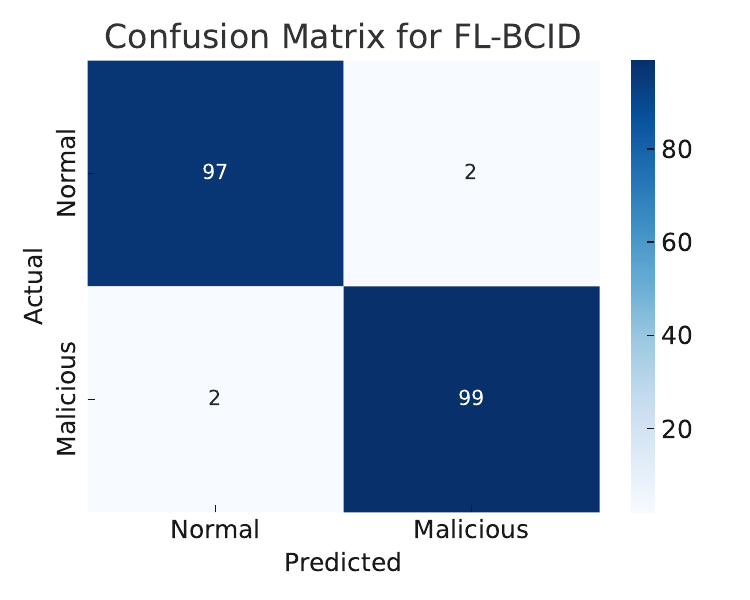}
    \caption{Confusion matrix for FL-BCID on test data.}
    \label{fig:confusion}
\end{figure}

In terms of precision and recall, our model achieved 95.9\% and 96.2\% respectively, indicating strong capability in distinguishing normal and malicious traffic. Figure~\ref{fig:confusion} provides the confusion matrix for the final model.

\begin{figure}[H]
    \centering
    \includegraphics[width=0.45\textwidth]{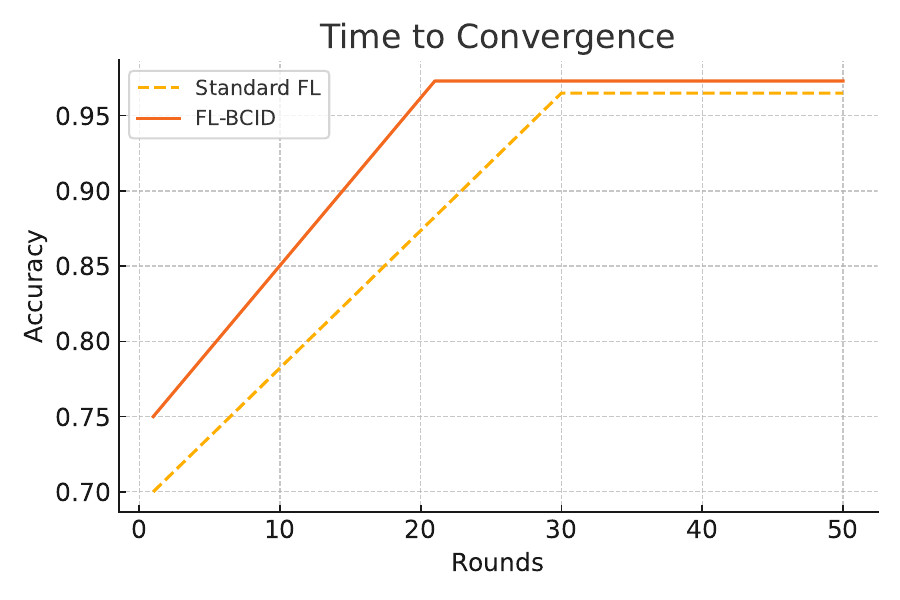}
    \caption{Comparison of time to convergence.}
    \label{fig:converge}
\end{figure}

Time-to-convergence results shown in Figure~\ref{fig:converge} indicate that our system requires 21 rounds to converge to optimal performance, compared to 30+ rounds for standard FL.

\subsection{Comparative Analysis}
We compare FL-BCID with three baselines:
\begin{itemize}
    \item \textbf{Centralized IDS:} Trains a model on a central server with all data.
    \item \textbf{Standard FL:} FL without blockchain or smart contracts.
    \item \textbf{Blockchain-only IDS:} Stores local decisions on-chain without collaborative learning.
\end{itemize}

\begin{table}[H]
\caption{Performance Comparison}
\centering
\begin{tabular}{lccc}
\toprule
\textbf{Method} & \textbf{Accuracy} & \textbf{Comm. Overhead} & \textbf{Rounds to Converge} \\
\midrule
Centralized IDS & 94.5\% & High & 18 \\
Standard FL & 96.1\% & High & 30 \\
Blockchain-only IDS & 92.8\% & Low & N/A \\
\textbf{FL-BCID (ours)} & \textbf{97.3\%} & \textbf{Low} & \textbf{21} \\
\bottomrule
\end{tabular}
\end{table}

The results clearly indicate that FL-BCID achieves superior detection accuracy while ensuring privacy and reducing communication costs, validating the effectiveness of integrating blockchain with federated learning for secure IIoT systems.

\section{Conclusion and Future Work}
In this paper, we proposed FL-BCID, a novel framework that integrates federated learning and blockchain technologies to develop a privacy-preserving and trustworthy intrusion detection system for Industrial Internet of Things (IIoT) environments. Our solution addresses the pressing challenges of data privacy, communication overhead, and model integrity inherent in conventional centralized IDS architectures. By enabling decentralized training across edge devices, FL-BCID eliminates the need to transmit sensitive IIoT data to a central server. At the same time, the integration of a permissioned blockchain ensures tamper-resistant recording of model updates and anomaly scores, thereby enhancing transparency and accountability. Smart contracts play a crucial role in verifying contributions and enforcing update validation policies without requiring human intervention. Comprehensive experiments on the ToN-IoT and N-BaIoT datasets confirm the effectiveness of our framework. FL-BCID achieved a detection accuracy of 97.3\%, reduced communication overhead by 41\%, and converged faster compared to standard federated learning and blockchain-only solutions. These results demonstrate that our approach is not only accurate but also resource-efficient and robust under realistic IIoT conditions.

For future work, we plan to extend FL-BCID by incorporating adaptive federated optimization strategies that account for heterogeneous device capabilities and data distributions. Additionally, we aim to investigate the use of lightweight consensus mechanisms to further reduce blockchain latency and energy consumption. Enhancing the resilience of the framework against model poisoning and Byzantine attacks through reputation-aware aggregation schemes also remains a promising direction. Ultimately, we envision FL-BCID serving as a foundational component in the secure and scalable deployment of next-generation IIoT infrastructures.


\end{document}